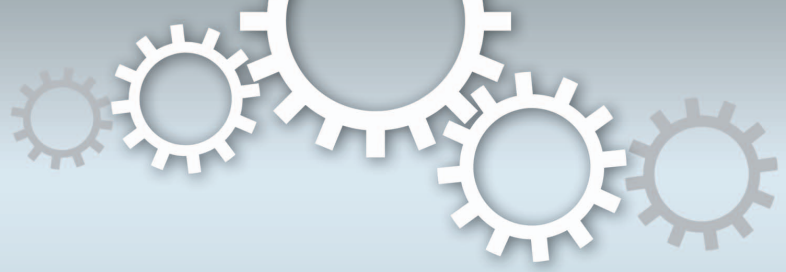

# SCIENTIFIC REPORTS

**OPEN**

# Field-induced quantum fluctuations in the heavy fermion superconductor CeCu$_2$Ge$_2$

D. K. Singh[1,2], A. Thamizhavel[3], J. W. Lynn[1], S. Dhar[3], J. Rodriguez-Rivera[1,2] & T. Herman[1]

[1]NIST Center for Neutron Research, Gaithersburg, MD 20899, USA, [2]Department of Materials Science and Engineering, University of Maryland, College Park, MD 20742, USA, [3]Tata Institute of Fundamental Research, Mumbai, INDIA.

Quantum-mechanical fluctuations in strongly correlated electron systems cause unconventional phenomena such as non-Fermi liquid behavior, and arguably high temperature superconductivity. Here we report the discovery of a field-tuned quantum critical phenomenon in stoichiometric CeCu$_2$Ge$_2$, a spin density wave ordered heavy fermion metal that exhibits unconventional superconductivity under $\simeq 10$ GPa of applied pressure. Our finding of the associated quantum critical spin fluctuations of the antiferromagnetic spin density wave order, dominating the local fluctuations due to single-site Kondo effect, provide new information about the underlying mechanism that can be important in understanding superconductivity in this novel compound.

E lectronic interactions in strongly correlated systems dictate the physical properties of quantum particles which ultimately results in different phases of matter.[1–3] Two such phases which often compete are super-conductivity and magnetism. However, in a set of materials, for example cuprates,[4] iron-based pnictides[5] and heavy fermions,[6] magnetism and superconductivity are not only found to coexist but a parallel resemblance in the energetic interplay between the two are also proposed.[7,8] Among the heavy fermion compounds of 4f- and 5f-electrons bands, Ce, Yb and U based, the underlying magnetism is controlled by the competing interaction between the single-ion Kondo effect and the long range Rudermann-Kittel-Kasuya-Yoshida (RKKY) exchange interaction.[9–11] The Kondo interaction hybridizes local moments with the conduction electrons to form Landau quasiparticles, whereas the RKKY interaction acts as the mediating interaction between these quasiparticles which often leads to the spin density wave (SDW) instability of the local moments of 4f- or 5f- electrons in a Fermi liquid of itinerant electrons.[12,13] In many cases, these competing magnetic interactions lead to a novel quantum critical ground state at $T = 0$ K, which separates the magnetic ordered state of $f$-electrons moments from the ensemble of weakly interacting local moments forming a paramagnetic state.[13–15] Since the Kondo and RKKY interactions reflect the local and long range antiferromagnetic behaviors, respectively, of the system, the microscopic nature of the critical fluctuations in the quantum critical state of a system can be dominated by either the local fluctuations or the fluctuation of antiferromagnetic order parameter. Local fluctuations due to the Kondo effect are single-site phenomena, thus wave-vector independent, as opposed to the fluctuations of long wave-length antiferromagnetic SDW order which are centered at the antiferromagnetic wave vectors. In order to understand the physical properties of this novel state, external parameters such as magnetic field or pressure or a tuning in the chemical composition are frequently used to drive the system towards the quantum critical state at $T \to 0$ K.[6]

In many heavy fermion compounds, the superconducting order parameter is found to emerge in the vicinity of a quantum critical point (QCP).[16–20] Therefore, a thorough understanding of the physical properties of this novel state is essential for a meaningful theoretical formulation. While the synergistic efforts of experimental and theoretical investigations have led to a broader understanding of this novel phenomenon, an unambiguous study of quantum critical behavior in a pure stoichiometric compound exhibiting the dual phenomena of quantum criticality and superconductivity is required. Here we report on the experimental investigation of the quantum magnetic properties in the stoichiometric compound CeCu$_2$Ge$_2$ which becomes superconducting, $T_c = 0.64$ K, under the pressure application of $\simeq 10$ GPa.[21] Our experimental aims are two-fold: (1) we want to explore the quantum phase transition (QPT) using magnetic field as the tuning parameter, and (2) clarify the nature of the driving mechanism behind QPT in a disorder free system to make the understanding conclusive. The detailed experimental investigation of CeCu$_2$Ge$_2$ using neutron scattering measurements confirm the presence of quantum fluctuations at low temperatures which leads to a quantum phase transition as $T \to 0$ K. We also show that the quantum critical phenomenon in this case is dominated by fluctuations of the AFM order parameter, well described by Hertz-Millis-Moriya (HMM) spin fluctuation theory which invokes the idea that the QPT results from the fluctuations of the AFM moment and their intensity diverges at the QCP.[22–25]

## Results

**Quantum fluctuations in CeCu₂Ge₂.** CeCu₂Ge₂ is an archetypal antiferromagnetic metal where long range magnetic order of the $Ce^{3+}$ ions develops below $T_N \simeq 4$ K.[26] The Ce moments in the $a^*$-$b^*$ plane form a spiral spin density wave (SSDW), illustrated in Figure 1a, with the propagation wave vector of $\mathbf{k} = (0.285, 0.285, 0.54)$. The plane of rotation of the spiral is perpendicular to $\mathbf{k}$. The magnitude and direction of the propagation vector is consistent with the theoretically determined wave vector connecting the parallel planes of the nested Fermi surface of CeCu₂Ge₂.[27] Theoretical calculations also suggest the existence of partial Kondo screening, below $T_K \simeq 6$ K, of the local moments by the weakly hybridized conduction bands. The itinerant character of the propagation vector is also exhibited in the temperature and field dependent measurements where small but continuous shifts in both $HH$- and $L$- components are observed as the parameters ($T$, $H$) are varied (see supplementary material).[28] The static moment associated with the long range incommensurate magnetic peaks, $\mathbf{Q_M}$, at $T = 1.5$ K is estimated to be $M_{Ce} \simeq 1.04(4)$ $\mu_B$, significantly smaller than that expected for a fully degenerate $J = 5/2$ ground state $Ce^{3+}$ ion value of 2.15 $\mu_B$. $\mathbf{Q_M}$ is related to the propagation wave vector $\mathbf{k}$ by $\mathbf{Q_M} = \mathbf{k} \pm \boldsymbol{\tau}$, where $\boldsymbol{\tau}$ represents structural Bragg peak position.

Interesting new results are found in the magnetic field measurements, applied perpendicular to the [HHL] scattering plane and thus parallel to the equivalent axes of (−110), which include the observation of (1) the disappearance of long range AFM order at $H_c \simeq 8$ T, and (2) near $H_c$, the static long range order is replaced by short range dynamic correlations which tend to become stronger as $T \to 0$ K; suggesting the quantum nature of the magnetic instability.[14] The $H - T$ phase diagram in Figure 1b describes the field-induced separation of the long-range ordered magnetic configuration from the short-range dynamic correlation across the $H_c$ boundary.

Figure 2a–f illustrate the two-dimensional maps of $\mathbf{q}$-scans of the dynamic structure factor $S(\mathbf{Q},E)$ at different energy transfers and measurement temperatures. At $H = H_c$, the magnetic Bragg peak at the incommensurate wave vector $\mathbf{Q_M}$ is destroyed and is replaced by broad features at finite energy transfers $E$. These short-range dynamic correlations are antiferromagnetic in nature in the field neighborhood of $H_c$. At $T = 0.2$ K, the intensity and the width of the dynamic structure factor first increase with increasing energy up to $\simeq 1$ meV then decrease rapidly to just above the background level, at $E = 1.5$ meV. We also note that the ratio between the spatial correlations along [110] and [001] directions at $E = 0.5$ meV is $\simeq 1.8$. Therefore CeCu₂Ge₂ is somewhat anisotropic. A small shift in the peak position of $S(\mathbf{Q},E)$ at different energy transfers, forming a *butter fly* pattern as shown in the inset of Fig. 2c, can also be seen, indicating the dispersive nature of the dynamic structure factor. Interestingly, the magnetic scattering at finite energy transfers occur in well correlated peaks around $\mathbf{Q_M}$. In order to establish the quantum nature of this short range dynamic behavior, we performed measurements at different temperatures. In the lower panel of Figure 2 the temperature dependences of $S(\mathbf{Q},E)$ at $E = 0.8$ meV are plotted, where we see that the intensity of the dynamic correlations reduces significantly as the measurement temperature is increased from $T = 0.2$ K to 2.8 K. This suggests that the fluctuations of the antiferromagnetically correlated Ce-spins are quantum fluctuations.

Next, we investigate the underlying mechanism behind these quantum fluctuations and explore the associated quantum phase transition. The characteristic signature of a quantum phase transition is manifested by the determination of the critical slowing down of the relaxation rate and is reflected in the energy linewidth of the excitation spectrum.[1,2] In the HMM formalism, the divergence of the static susceptibility as $T \to 0$ K is also expected.[22] Alternatively, if the quantum critical behavior is dominated by the *local* Kondo interaction, the static susceptibility follows the Curie-Weiss ansatz.[9,13] In order to determine the linewidth of the excitation spectrum, measurements were performed as a function of energy at the magnetic wave-vecor $\mathbf{Q_M} = (0.29, 0.29, -0.54)$ for various temperatures between $T = 0.2$ K and 40 K. Measurements were also performed at another wave vector $\mathbf{Q_0} = (0.38, 0.38, -0.2)$, sufficiently far away from $\mathbf{Q_M}$, where the inelastic spectra can be considered due to the local magnetic fluctuations arising from the single-site Kondo interaction. Figure 3a–b show the background corrected representative scans at $\mathbf{Q_M}$ and $\mathbf{Q_0}$ at few selected temperatures, above and below $T_N$, and at $H \simeq 8$ T. Plots in Figure 3a confirm the qualitative behavior observed in the temperature dependence of $\mathbf{q}$-scans in Figure 2, where a strong enhancement in the antiferromagnetic fluctuations is observed at lower temperatures. In Figure 3b, we see little variation in the intensity of the local fluctuations at $\mathbf{Q_0}$ with $T$. At high temperatures, the spin fluctuations at the AFM wave vectors cross into the thermal regime.

The quantitative determinations of the relaxation rate and the static susceptibility as a function of temperature are performed using standard neutron scattering intensity analysis,[16,29] which can be described by,

$$S(\mathbf{Q}, E, T) = \frac{1}{\pi} \frac{1}{1 - e^{-E/kT}} \chi''(\mathbf{Q}, E, T) \qquad (1)$$

where $\chi''$ is the dynamic spin susceptibility. The dynamic

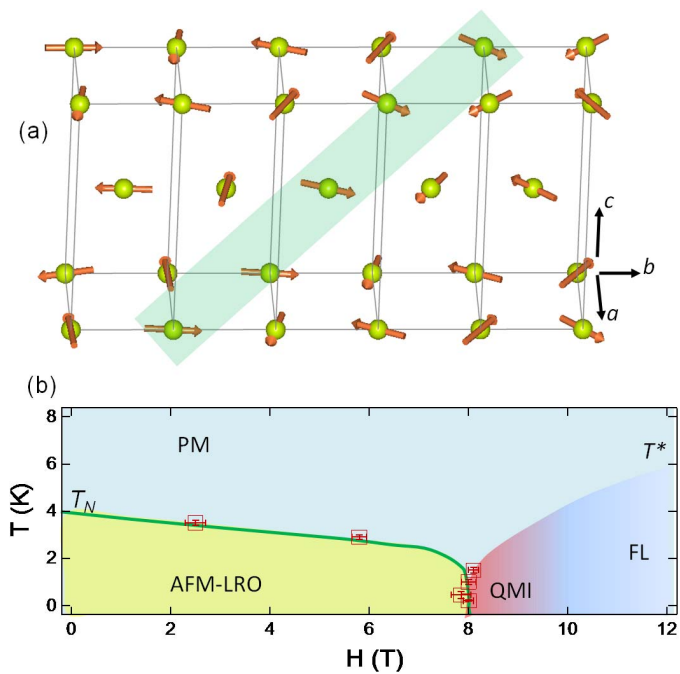

**Figure 1 | Spin structure of correlated Ce ions and magnetic phase diagram of CeCu₂Ge₂.** (a), Correlated Ce moments, in the $a$–$b$ plane, form a spiral spin density wave configuration. The shaded area highlights the spiral plane of rotation perpendicular to the propagation vector $\mathbf{k}$. (b), $H$–$T$ magnetic phase diagram of CeCu₂Ge₂ exhibiting various regimes of antiferromagnetic long range order (AFM-LRO), quantum magnetic instability (QMI), Fermi liquid and paramagnetic behaviors. $T_N$ and $T^*$ are the Neel AFM order and Kondo screening temperatures, respectively.

                    2

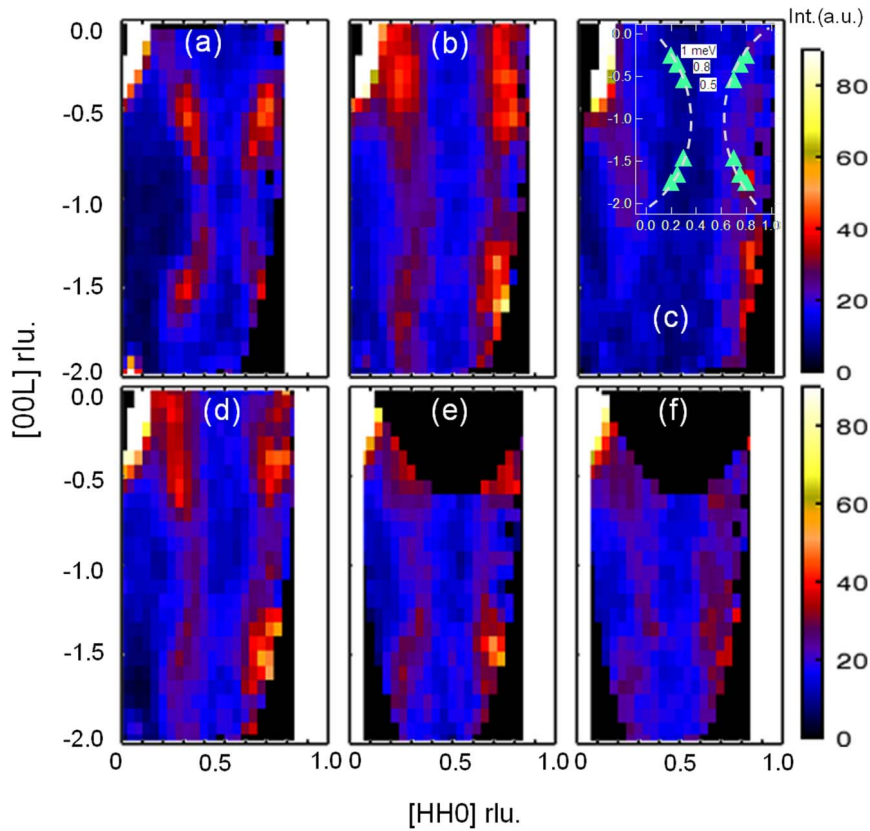

**Figure 2 | q-scan maps at different energies and temperatures at H = H_c.** (a–c), q-scans in reciprocal space at different energy transfers of 0.5, 1 and 1.5 meV, respectively, at H = 8 T and T = 0.2 K. Short-range dynamic correlation of the Ce-ions are clearly observed at low energies. Modest change in peak positions at different energy transfers, forming the butterfly pattern along the dashed white curves (as shown in the inset of **c**), indicate the dispersive nature of the fluctuations. (d–f), Measurements at temperatures of 0.2, 1.5 and 2.8 K, respectively, at an energy transfer of E = 0.8 meV and H = H_c. Short-range dynamic correlations become weaker as the temperature is increased, indicating the quantum nature of the fluctuations.

susceptibility is described by the Fourier transform of the exponential decay of relaxation rate $\Gamma(\mathbf{Q},T)$ and is related to the total and static susceptibilities $\chi$ and $\chi'$, respectively, via the Kramers-Kronig relation. This is given by:

$$\chi(\mathbf{Q}, E, T) = \chi'(\mathbf{Q}, E, T) + i\chi''(\mathbf{Q}, E, T) \quad (2)$$

$$= \frac{A(\mathbf{Q},T)}{\Gamma(\mathbf{Q},T) - iE} \quad (3)$$

$$= \frac{\chi'(\mathbf{Q},T) \times \Gamma(\mathbf{Q},T)}{\Gamma(\mathbf{Q},T) - iE} \quad (4)$$

$\chi'(\mathbf{Q},T)$ was determined by integrating $\chi''(\mathbf{Q}, E, T)/E$ over the experimental energy range between 0 and 3 meV, as the energy spectra of fluctuations in this case are limited to low energy.[29] Inelastic neutron scattering data in Fig. 3 are well fitted using the above equations, giving a single Lorentzian lineshape. The extracted values of $\chi'(\mathbf{Q},T)$ and $2\Gamma(\mathbf{Q},T)$ (full width at half maximum, FWHM) at both $\mathbf{Q}_M$ and $\mathbf{Q}_0$ are plotted as a function of temperature in Figure 3e and 3f. In these plots, we see that a continuous increase in the static susceptibility is accompanied by a continuous decrease in the linewidth at $\mathbf{Q}_M$ as the temperature decreases. The decrease in $\Gamma(\mathbf{Q})$ becomes faster as the system passes through the zero-field AFM transition at $\simeq 4$ K and critically slows down as $T\to 0$ K. Also noticeable is the diverging behavior in $\chi'(T)$ as the temperature is reduced to 0.2 K. At $\mathbf{Q}_0$, on the other hand, no such behavior is observed. Finite but almost temperature independent (within error bar) $\chi'(\mathbf{Q},T)$ and $\Gamma(\mathbf{Q},T)$ at $\mathbf{Q}_0$ confirm the presence of a weak local Kondo-type fluctuations in the antiferromagnetic phase. The energy linewidth of the inelastic spectra at $\mathbf{Q}_0$ tends to saturate around

0.6 meV, which can be identified with the characteristic temperature of the Kondo screening interaction, $T^* \simeq 6.5$ K. The parallel presence of Kondo screening with a localized antiferromagnetic interaction poses an important question: are the bulk properties dominated by the local Kondo screening term or the magnetic wave-vector dependent fluctuations of the AFM order parameter. To answer this question, we performed bulk susceptibility measurements for H parallel to the same ($-110$) direction.

Bulk susceptibility data that are dominated by the Kondo screening interaction are expected to follow the Curie-Weiss (CW) rule.[2, 9] As shown in Figure 4, we observe that (1) the inverse susceptibility curves at low fields show a sharp upturn at $T_N$, indicating the transition to the AFM ordered state and (2) near the critical field at $H = 8.5$ T, the sharp upturn at $T \simeq 4$ K is replaced by a curve, significantly deviated from the CW line, followed by a weak upturn below $T \simeq 2.8$ K. These measurements suggest that the AFM correlated order parameter fluctuations dominate the local fluctuations due to the Kondo effect. As the applied field is increased beyond $H_c (\simeq 8$ T) such that the magnetic system moves further away from the quantum critical regime, the effect of correlated AFM fluctuations starts fading and the local Kondo interaction starts dominating.

**HMM-type scaling of the dynamic spin susceptibilities.** The analysis of the experimental data clearly supports the existence of field-induced quantum fluctuations, leading to a quantum phase transition as $T \to 0$ K. The combination of neutron scattering and bulk susceptibility measurements strongly support the scenario where the AFM order parameter fluctuations at $\mathbf{Q}_M$ are responsible for the field-induced quantum phase transition in CeCu$_2$Ge$_2$ and dominate the critical properties of the system.[30, 31]

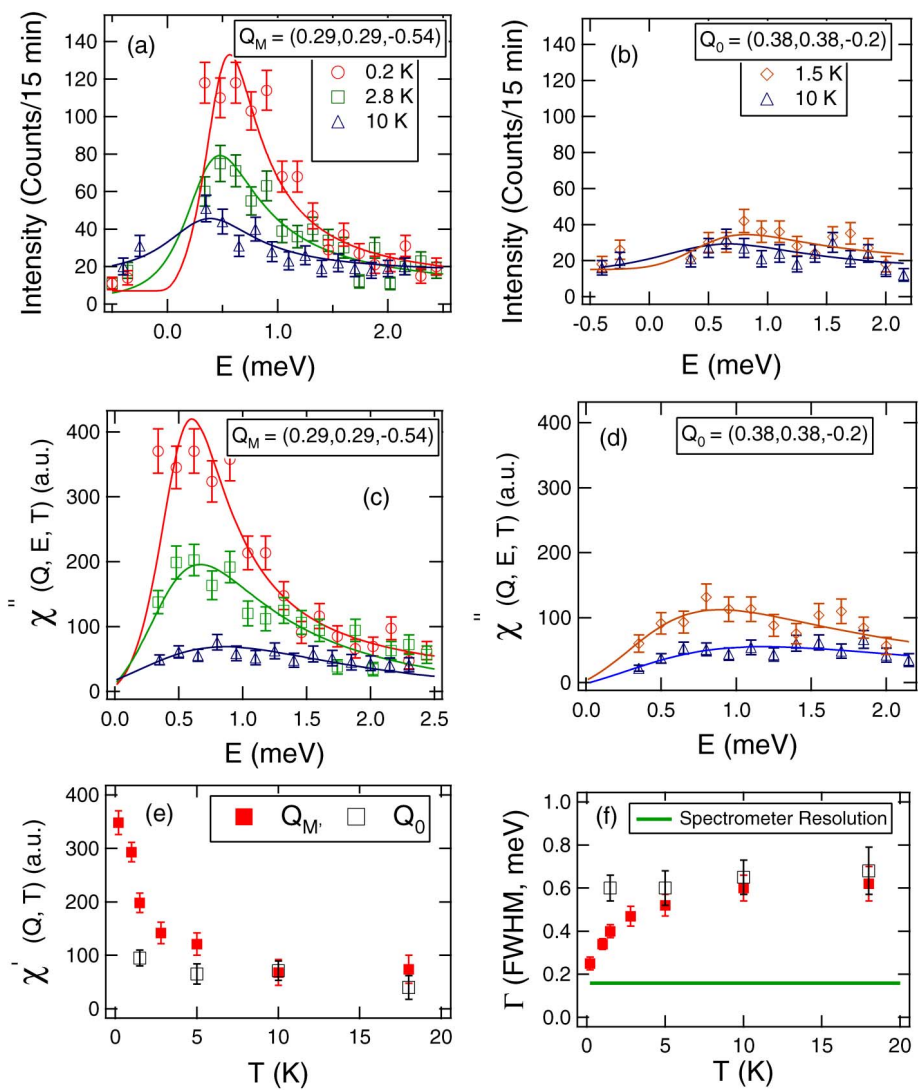

**Figure 3 | Inelastic neutron scattering measurements.** (a–b), Representative scans of inelastic measurements for the magnetic correlations at, $Q_M$, and the local Kondo fluctuations $Q_0$ at different temperatures and $H \simeq 8$ T. Solid lines are Lorenztian fits to the data, as explained in the text. Error bars represent one standard deviation. (c–d)Dynamic spin susceptibilities at $Q_M$ and $Q_0$ are extracted from the inelastic neutron scattering data. (e), Temperature dependence of the static susceptibility at $Q_M$ and $Q_0$. The static susceptibility at $Q_M$ tends to diverge as $T{\to}0$ K. (f), Temperature variation of the relaxation rate, $2\Gamma(Q,T)$ (meV), at both $Q$s at $H = 8$ T. Green line indicates the Spectrometer resolution. Clearly, the long wavelength spin fluctuations critically slow down at $Q_M$ as $T{\to}0$ K.

This argument can be further validated by performing a scaling analysis of the dynamic spin susceptibility data in the form $\chi'.T^\alpha = f(E/T^\beta)$,[9, 29] It is argued that the Curie-Weiss behavior reflected in the local-moment fluctuations leads to a linear ($E/T$) scaling.[13] On the other hand, if the spins fluctuations are dominated by the AFM order parameter then $\beta$ is 1.5.[24, 32, 33] Scaling plots of dynamic susceptibilities, obtained using equation 1, are plotted in Figure 5 for two different sets of parameters: (a) $\alpha = 1.5$, $\beta = 1.5$ and (b) $\alpha = 0.8$, $\beta = 1$. Clearly, the SDW-type scenario gives a much better collapse of the data onto one curve compared to the locally critical scenario. Experimental data in (a) is fitted using the HMM formulation leading to $f(x) = \frac{abx}{1+(bx)^2}$. The scaling of the dynamic susceptibilities using the HMM formulation further corroborates our view that the field-induced quantum fluctuations in $CeCu_2Ge_2$ are primarily of the long wavelength antiferromagnetic origin.

## Discussion

Our experimental investigation of the stoichiometric compound $CeCu_2Ge_2$ reveals the presence and nature of the quantum critical behavior, whch were unknown in this archetypal system due to the technical difficulties associated with accessing the high pressure experimental regime. Our approach of using field as the external tuning agent, instead of pressure, demonstrates that the quantum critical phenomenon is intrinsic to the system, like superconductivity, and can be independently explored. These two novel behaviors are very likely directly related,[6, 20] but detailed calculations of the energetics of the system in both the superconducting and the quantum critical states are required to establish this proposition. In recent work on the isoelectronic sister compound $CeCu_2Si_2$, the role of the field-induced quantum critical state as the driving agent behind superconductivity was established via the energetics comparison in the two states.[16] Another important finding from our study on $CeCu_2Ge_2$ is the identification of the fluctuations of the AFM order parameter, well described by the HMM spin fluctuation theory, as the underlying mechanism behind the quantum critical phenomenon. Recently, there has been considerable discussion about the applicability of the HMM formulation in explaining the quantum critical phenomenon.[12, 34–36] In this regard, our results from the study of a stoichiometric compound certainly sheds new light, as the stoichiometric nature of the system rules out any effect due to chemical disorder. Future experimental research involving the change in the Fermi surface properties near the QCP are highly desirable to further understand this phenomenon in this compound.

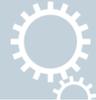

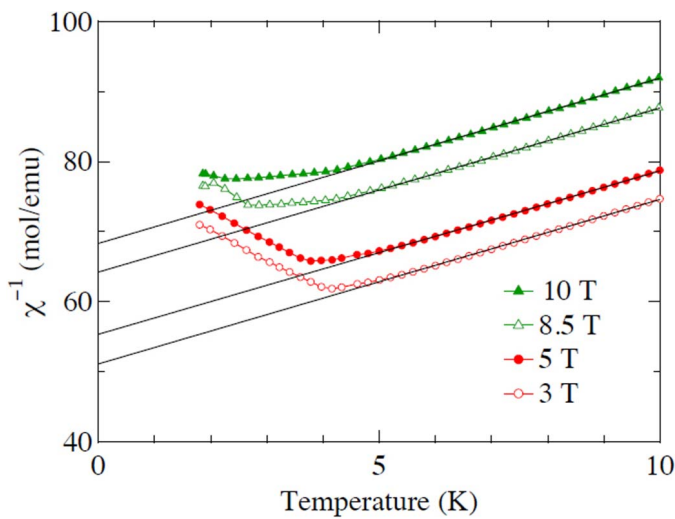

**Figure 4 | Bulk inverse susceptibility measurements of CeCu₂Ge₂** Measurements in identical field alignment show a significant deviation from Curie-Weiss local behavior at low temperatures. The inverse susceptibility plots at different fields are vertically offset by finite values for clarity. As the applied field is increased, the system moves further away from the quantum magnetic instability regime and tends to gradually recover Fermi-liquid properties. This fact is reflected by the diminishing deviation of the inverse susceptibility from the CW line at higher fields.

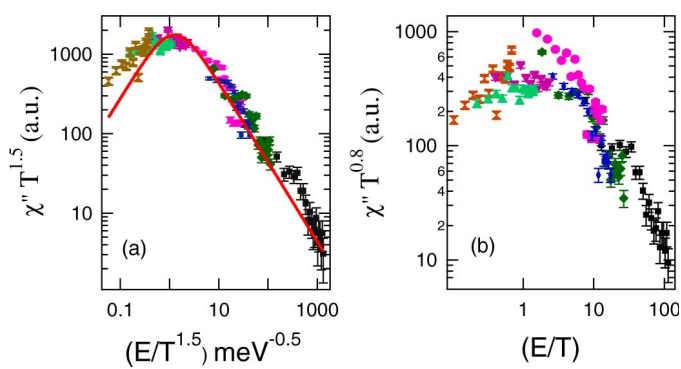

**Figure 5 | Scaling plots of the dynamic susceptibility data.** (a), HMM scenario. (b), Local scenario. The dynamic susceptibility data at different temperatures correlate well in HMM scenario while a very poor correlation is observed in the *local* scenario, further suggesting that the quantum critical behavior is dominated by the fluctuation of AFM order parameter.

## Methods

Measurements were performed on high quality single crystals, grown by the flux method. Characterization of the crystals were performed using X-ray measurements, confirming the high quality of the crystals. The inelastic neutron scattering measurements were performed on one rectangular shape single crystal with a mass of 1.8 gm on the SPINS and MACS cold spectrometers at the NIST Center for Neutron Research with fixed final energy of 3.5 meV and cooled He filters after the sample. At this fixed final energy, the spectrometers resolution (FWHM) were determined to be $\simeq$ 0.15 and 0.2 meV, respectively. The horizontal collimations were 32'-80'-Sample-80'-120'. The sample was mounted in a superconducting vertical field magnet in the [HHL] scattering plane with $a = b = 4.345$ Å and $c = 10.94$ Å. Perpendicular field application direction was ($-110$) direction in this tetragonal system. Inelastic measurements for quantitative analysis were performed at SPINS while the q-scan maps were obtained at MACS. Bulk susceptibility measurements were performed on a separate single crystal, flux grown in the same batch, under identical experimental conditions of field alignment.

1.  Coleman, P. and Schofield, A. Quantum criticality. *Nature* **433**, 226–229 (2005).
2.  Sachdev, S. Quantum phase transitions. (Cambridge Univ. Press, New York, 1999).
3.  Varma, C. M. Mixed-valence compounds. *Rev. Mod. Phys.* **48**, 219–238 (1976).
4.  Tranquada, J. M. *et al.* Quantum magnetic excitations from stripes in copper oxide superconductors. *Nature* **429**, 531–534 (2004).
5.  Cruz, C. de la *et al.* Magnetic order close to superconductivity in the iron-based layered LaO₁₋ₓFₓFeAs. *Nature* **453**, 899–902 (2008).
6.  Gegenwart, P. *et al.* Quantum criticality in heavy fermion metals. *Nature Physics* **4**, 186 (2008).
7.  Fisk, Z. *et al.* Heavy fermion metals: New highly correlated states of matter. *Science* **239**, 33 (1988)
8.  Mathur, N. D. *et al.* Magnetically mediated superconductivity in heavy fermion compounds. *Nature* **394**, 39 (1998).
9.  Schroder, A. *et al.* Onset of antiferromagnetism in heavy fermion metals. *Nature* **407**, 351 (2000).
10. Custers, J. *et al.* The break-up of heavy electrons at a quantum critical point. *Nature* **424**, 524 (2003).
11. Aronson, M. C. *et al.* Non-Fermi-liquid scaling of the magnetic response in UCu₅₋ₓPdₓ. *Phys. Rev. Lett.* **75**, 725–728 (1995).
12. Coleman, P. *et al.* How do Fermi liquids get heavy and die. *J. Phys.: Cond. Matt.* **13**, R723 (2001).
13. Si, Q. *et al.* Locally critical quantum phase transitions in strongly correlated metals. *Nature* **413**, 804 (2001).
14. Hewson, A. C. *The Kondo problem to Heavy Fermions.* (Cambridge Univ. Press, Cambridge, 1993).
15. Lohneysen, H. *et al.* Fermi-liquid instabilities at magnetic quantum phase transitions. *Rev. Mod. Phys.* **79**, 1015 (2007).
16. Stockert, O. *et al.* Magnetically driven superconductivity in CeCu₂Si₂. *Nature Physics* **7**, 119 (2011).
17. Ebihara, T. *et al.* Emergent fluctuation hot spots on the Fermi surface of CeIn₃ in strong magnetic fields. *Phys. Rev. Lett.* **93**, 246401 (2006).
18. Sato, N. K. *et al.* Strong coupling between local moments and superconducting heavy electrons in UPd₂Al₃. *Nature* **410**, 340 (2001).
19. Saxena, S. S. *et al.* Superconductivity on the border of itinerant-electron ferromagnetism in UGe₂. *Nature* **406**, 587 (2000).
20. Stewart, G. R. *et al.* Possibility of coexistence of bulk superconductivity and spin fluctuations in UPt₃. *Phys. Rev. Lett.* **52**, 679–682 (1984).
21. Jaccard, D. *et al.* Pressure-induced heavy fermion superconductivity in CeCu₂Ge₂. *Physics Letters A* **163**, 475 (1992).
22. Knafo, W. *et al.* Antiferromagnetic criticality at a heavy-fermion quantum phase transition. *Nature Physics* **5**, 753 (2009).
23. Hertz, J. A. Qunatum critical phenomena. *Phys. Rev. B* **14**, 1165 (1976).
24. Millis, A. J. Effect of a nonzero temperature on quantum critical points in itinerant fermion systems. *Phys. Rev. B* **48**, 7183 (1993).
25. Moriya, T. *et al.* Anomalous properties around magnetic instability in heavy electron systems. *J. Phys. Soc. Jpn* **64**, 960 (1995).
26. Knopp, G. *et al.* Magnetic order in a Kondo lattice: a neutron scattering study of CeCu₂Ge₂. *Z. Phys. B* **77**, 95 (1989).
27. Zwicknagl, G. Kondo effect and antiferromagnetism in CeCu₂Ge₂: An electronic structure study. *J. Low Temp. Phys.* **147**, 123 (2007).
28. Krimmel, A. *et al.* Single crystal neutron diffraction studies on CeCu₂Ge₂ and CeCu₁.₉Ni₀.₁Ge₂. *Phys. Rev. B* **55**, 6416 (1997).
29. Knafo, W. *et al.* Anomalous scaling behavior of the dynamic spin susceptibility of Ce₀.₉₂₅La₀.₀₇₅Ru₂Si₂. *Phys. Rev. B* **70**, 174401 (2004).
30. Moriya, T. *Spin Fluctuations in Itinerant Electron Magnetism* (Springer, 1985).
31. Lonzarich, G. *et al.* Spin density fluctuations in magnetic metals. *Physica B* **156–157**, 699–705 (1989).
32. Stockert, O. *et al.* Magnetic fluctuations at a field-induced quantum phase transition. *Phys. Rev. Lett.* **99**, 237203 (2007).
33. Continentino, M. A. *Quantum Scaling in Many-Body Systems* (World Scientific, 2001).
34. Anderson, P. W. *et al.* A Fermi sea of heavy electrons (a Kondo lattice) is never a Fermi liquid. Preprint at http://arxiv.org/abs/0810.0279v1 (2008)
35. Saremi, S. *et al.* Unifying Kondo coherence and antiferromagnetic ordering in the honeycomb lattice. *Phys Rev B* **83**, 125120 (2011).
36. Zhu, J.-X. *et al.* Continuous quantum phase transition in a Kondo lattice model. *Phys. Rev. Lett.* **91**, 156404 (2003).

## Acknowledgements

This work was financially supported in part by the NSF under Agreement No. DMR-0944772. We thank Q. Stock, Y. Zhao and S. Chang for helping with the neutron scattering measurements.

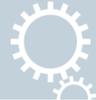

## Author contributions

D.K.S and A.T. contributed equally to this work. The crystals were grown by A.T., the experiments were carried out by D.K.S., A.T., S.D., J.R. and T.H., the analysis were carried out by D.K.S., A.T. and the paper was written by D.K.S. and J.W.L.

## Additional information